# Space Weather

## Interplanetary Magnetic Field Control of Polar Ionospheric Equivalent Current System Modes

R. M. Shore[1] 🔟, M. P. Freeman[1] 🔟, and J. W. Gjerloev[2,3] 🔟

[1]British Antarctic Survey, Cambridge, UK, [2]Johns Hopkins University, Applied Physics Laboratory, Laurel, MD, USA, [3]Birkeland Center of Excellence, Department of Physics and Technology, University of Bergen, Bergen, Norway

**Abstract** We analyze the response of different ionospheric equivalent current modes to variations in the interplanetary magnetic field (IMF) components $B_y$ and $B_z$. Each mode comprises a fixed spatial pattern whose amplitude varies in time, identified by a month-by-month empirical orthogonal function separation of surface measured magnetic field variance. Here we focus on four sets of modes that have been previously identified as DPY, DP2, NBZ, and DP1. We derive the cross-correlation function of each mode set with either IMF $B_y$ or $B_z$ for lags ranging from −10 to +600 mins with respect to the IMF state at the bow shock nose. For all four sets of modes, the average correlation can be reproduced by a sum of up to three linear responses to the IMF component, each centered on a different lag. These are interpreted as the statistical ionospheric responses to magnetopause merging (15- to 20-min lag) and magnetotail reconnection (60-min lag) and to IMF persistence. Of the mode sets, NBZ and DPY are the most predictable from a given IMF component, with DP1 (the substorm component) the least predictable. The proportion of mode variability explained by the IMF increases for the longer lags, thought to indicate conductivity feedbacks from substorms. In summary, we confirm the postulated physical basis of these modes and quantify their multiple reconfiguration timescales.

## 1. Introduction

Electrodynamical coupling between the magnetosphere and the interplanetary magnetic field (IMF) drives current systems which span near Earth space and which are highly variable on a wide range of spatial and temporal scales (Dungey, 1961; Schunk & Nagy, 2009). The ionospheric footprints of these current systems generate noisy magnetic fields, which negatively impact efforts to model the Earth's internal magnetic field (Finlay et al., 2016; Thébault et al., 2017), and create geomagnetically induced currents which damage power grids (Beggan et al., 2013). It is desirable to better understand and predict this interconnected system in order to help mitigate these impacts.

Unfortunately, at present, it is only possible to predict the ionospheric currents (or their associated magnetic perturbation at ground) with rather low levels of accuracy from measurements of the IMF, particularly on small spatial scales. This is because while the ionospheric current systems are ultimately driven by disturbances happening on the Sun, these systems vary in the extent to which they are driven either directly by the IMF or by internal magnetospheric processes, and the entire set of current systems is subject to feedbacks operating within and between the ionosphere and magnetosphere. These feedbacks include the inertia of neutral winds, changes in ionospheric conductivity from particle precipitation, and the acceleration of electrons by parallel electric fields in the upper ionosphere caused by intense field-aligned currents (Knight & Parallel electric fields, 1973).

Despite this complexity, it has long since been shown (e.g., Nishida, 1966, 1968b; Obayashi & Nishida, 1968) that the morphology of the external magnetic perturbations comprises a set of large-scale ionospheric equivalent current systems, each driven by specific components of the IMF. For instance, the two-cell ionospheric convection vortices (termed Disturbance Polar type 2 or DP2) are strongly driven by the negative (southward) IMF $B_z$ component (Friis-Christensen & Wilhjelm, 1975; Hairston et al., 2005; Nishida, 1968a). Positive (northward) IMF $B_z$ fluctuations are associated with the NBZ system (Friis-Christensen et al., 1985; Maezawa, 1976). The IMF $B_y$ (dawn-dusk) component drives the DPY equivalent current system (Friis-Christensen & Wilhjelm, 1975; Friis-Christensen et al., 1985). In addition to these directly driven components, there is also the DP1 system, which is associated with the substorm current wedge and hence is indirectly driven by the IMF (e.g., Morley et al., 2007).

Historically, these DP systems have been identified from many analyses, each based on small amounts of data. Now, we have access to large quantities of uniform quality data, allowing us to investigate the ionosphere in a more objective manner. Here we use the model of the surface external and induced magnetic field (SEIMF) produced by Shore et al. (2018). The SEIMF model spans 12 years from 1997.0 to 2009.0 and uses the method of Empirical Orthogonal Functions (EOF) to isolate and identify the individual polar ionospheric equivalent current systems. The method, which minimizes the need for assumptions of physical behavior when assessing large quantities of geophysical data, is described in more detail in the following section.

The focus of our study is to investigate the nature of the IMF forcing of the ionosphere through its effects on the geomagnetic field variation. We focus on both the time delay of the ionospheric response and the proportion of the ionospheric equivalent current system variability described by the IMF. This allows us to also systematically test the hypothesis of Shore et al. (2018) that the dominant patterns of large-scale variability in the ionospheric equivalent currents are describable as the historic disturbance-polar systems. The delays in the effects of given IMF components on measurements or indices of the magnetosphere-ionosphere system have been studied previously, for example, by Nishida and Maezawa (1971), Meng et al. (1973), Baker et al. (1981), Bargatze et al. (1985), Browett et al. (2017), and Maggiolo et al. (2017). Our study is the first to systematically investigate the individual responses of the DP2, DPY, NBZ, and DP1 systems to IMF driving over the scale of a solar cycle.

In section 2, we describe the data we use to represent the DP2, DPY, NBZ, and DP1 equivalent current systems and the IMF $B_y$ and $B_z$ components; in section 3, we describe our method of comparing the response of the equivalent current systems to the IMF, and in section 4, we present the results of these comparisons. The significance of our findings are discussed in section 5, and we summarize in section 6.

## 2. Data

The data we use to represent the individual polar ionospheric equivalent current systems are the spatial and temporal SEIMF patterns isolated and identified by Shore et al. (2018). The authors applied the EOF method to individual months of ground-sampled vector magnetic field data from the SuperMAG archive (Gjerloev, 2012). The EOF method decomposes these data into spatiotemporal basis vectors (i.e., patterns defined in space and time) which collectively describe the majority of the variance of the data. For a given month of magnetic field data used as input to the EOF method, the output of the EOF decomposition is a series of independent spatial patterns, each of which has an associated amplitude variation in time (with 5-min resolution). A paired spatial and temporal pattern is one basis vector of the EOF decomposition, ranked by its contribution to the total variance. Shore et al. applied graph theory (Caldarelli, 2007) to the spatial patterns of the leading six modes in order to identify spatially similar clusters in the basis vectors from 144 sequential monthly sets of EOF analyses spanning the 12 years from 1997.0 to 2009.0. These spatiotemporal clusters were considered to be individually dominated by specific equivalent current systems including DP2, DPY, DP1, and NBZ. Here we test the hypothesis that the clusters of EOF patterns indeed represent those equivalent current systems. We use the Shore et al. model to provide a set of consecutive time series describing the amplitudes of these equivalent current systems, spanning 144 months across solar cycle 23. The DP2 set spans the full 144 months, while the other sets are not temporally complete: DPY spans 42 of the full 144 months, DP1 spans 70 months, and NBZ spans 10 months. The reason for this incompleteness is either that the equivalent current system amplitude is too weak in some months (e.g., DPY is not identified in winter) or that the equivalent current system's representation in the EOF patterns is not spatially coherent enough between all months to be fully described by the graph theory process.

Within each month, the product of an EOF spatial pattern and its associated temporal series gives a prediction of the SEIMF variations of a given equivalent current system. Thus, each EOF pattern allows us to represent a (known) large-scale two-dimensional spatial distribution of vector magnetic perturbation via its associated 1-D time series of amplitudes.

Starting from the assumption that a perturbation in a given IMF component will produce some (unknown) spatial perturbation in the ionospheric equivalent currents, we can use the temporal correlation between the IMF measurements and the EOF time series (associated with a known spatial pattern) to probe the extent to which the IMF-driven perturbations have that specific spatial form, over a given timespan. This span is a trade-off between obtaining a large enough range of solar wind conditions and external field variability while avoiding the introduction of variance from other factors (e.g., seasonal differences). Here we choose the timespan to be 1 month.

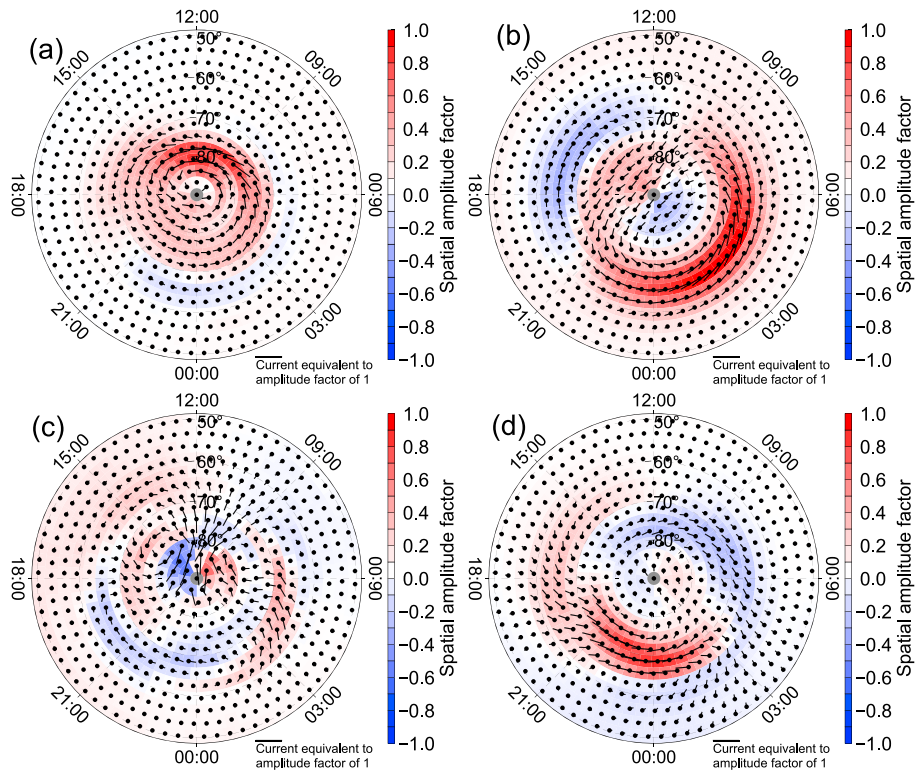

**Figure 1.** Mean spatial patterns for the equivalent current systems (a) DPY, (b) DP2, (c) NBZ, and (d) DP1. For each pattern, the mean is taken across all spatial patterns from each epoch, for which there is a representation of that equivalent current system in the data set of Shore et al. (2018). The background colored values are those of the Quasi-Dipole magnetic $\theta$ component, and the vectors are the horizontal magnetic component rotated by 90° clockwise to indicate the direction and relative strength of the equivalent currents (assuming an even distribution of ionospheric conductivity). The temporal perturbations assessed throughout this study pertain to the sense of these patterns. The polar bin has no coverage and is shown in gray.

Since the EOF patterns are those which maximize variance, this temporal correlation with the IMF is to be expected to be reasonably good, and indeed, this has been demonstrated by Shore et al. (2017) and Shore et al. (2018). The authors also showed that the different EOF patterns—and the individual ionospheric equivalent current systems they were considered to represent—have well-separated characteristic correlations with the different IMF components. Hence, this temporal correlation assessment can tell us the spatial and temporal nature of the coupling of individual IMF components with the dominant component systems of the ionospheric electrodynamics.

To represent the IMF variations, we use 1-min measurements of the IMF strength in the GSM $y$ and $z$ directions—$B_y$ and $B_z$ (Hapgood, 1992)—from the National Aeronautics and Space Administration Advanced Composition Explorer (ACE) spacecraft (Stone et al., 1998) lagged from near the L1 Lagrangian point to their arrival time at the bow shock nose. In the correlation analysis presented here, we apply additional lags to the IMF measurements ranging from −10 to +600 min (in 5-min steps up to 150 min and then 10-min steps thereafter), where a positive lag of $\tau$ minutes means that we expect a disturbance measured at the bow shock to take a further $\tau$ minutes to have an impact in magnetic records taken at the Earth's surface. Following the lagging process, we take 5-min means of the lagged IMF measurements, corresponding to the epochs of the SEIMF time series within each month. The physical significance of applying these lags is that we expect the peak temporal correlation to occur at the lag time at which the (unknown) IMF spatial perturbation best matches the spatial form of the EOF pattern. Thus, we are assessing the reconfiguration timescale of the associated equivalent current system.

The spatial and temporal patterns which comprise a given EOF basis vector are sign indeterminate. The basis vector—which is the product of the spatial and temporal patterns—will always have the same overall sign if we were to repeat the EOF analysis of a given data set. Yet the component spatial and temporal parts of that basis vector could individually vary in sign between repeated decompositions of the same data set. Since we will use the EOF time series here in isolation from their associated spatial patterns, we must force them to have a consistent sign across all months. After which, their correlations with the solar wind measurements will be comparable between different monthly epochs. We do this for each equivalent current system as follows. We take the temporal pattern from a given reference month and correlate the temporal patterns (for the same equivalent current system) from all other months with it. For each $j$th month, if the correlation with the reference month is negative, we reverse the sign of the $j$th time series, and we also reverse the sign of the associated $j$th spatial pattern. Following this, the temporal and spatial patterns for each equivalent current system will have the same relative sense between all individual months. The mean of the sign-corrected spatial patterns for the equivalent current systems DPY, DP2, NBZ, and DP1 is shown in Figure 1. These patterns are the same as those shown in Shore et al. (2018), but their sense (sign) here represents the spatial distribution of equivalent current for a unit positive temporal perturbation. These relative signs are used throughout the remainder of this report.

## 3. Method

We compute the linear Pearson correlation (Press, 1992) between the time series of each equivalent current system and the (lagged and resampled) IMF $B_y$ and $B_z$ series, once per lag. The SEIMF models each span one calendar month plus 1 day either side of the month, and thus, each correlation computation is nominally based on between 8,640 and 9,504 temporal data point pairs (at 5-min resolution). The patchy coverage in the ACE measurements will vary the true data count since we ignore epochs where ACE data are missing (there are no missing temporal data in the SEIMF models).

To contextualize these cross correlations of the solar wind driver and the ionospheric response, we model the response of each equivalent current system with a scaled, time-shifted version of the autocorrelation of the driver, as follows. Consider the SEIMF variation $Y(t)$, given by 1-month-long time series of an EOF mode amplitude, and an interplanetary driver $X(t)$, given by the corresponding IMF $B_y$ or $B_z$ time series. Now let us assume that the SEIMF response is in part directly driven by the solar wind driver such that we can write

$$Y(t) = aX(t - \tau_a) + W(t),$$  (1)

where $a$ measures the strength of the response to the driver, $\tau_a > 0$ is the delay in the response with respect to the driver, and $W$ is the part of the response that is unrelated to the driver.

The mean of $Y$ is given by

$$
\begin{aligned}
\mu_Y &= E[Y] \\
&= E[aX(t - \tau_a) + W(t)] \\
&= aE[X(t - \tau_a)] + E[W(t)] \\
&= aE[X(t)] + E[W(t)] \\
&= a\mu_X + \mu_W,
\end{aligned}
$$  (2)

where $E[]$ denotes the expected value of the quantity in brackets (Ross, 2006), and we have assumed that $X(t)$ is stationary (i.e., we assume that the mean over a month span is insensitive to shifts in this span of up to the maximum lag, which is 10 hr).

Now the Pearson cross-correlation coefficient between $Y(t)$ and $X(t - \tau)$ is given by

$$r_{XY}(\tau) = \frac{C_{XY}(\tau)}{\sigma_X \sigma_Y},$$  (3)

where $\sigma_X$ and $\sigma_X$ are the standard deviations of $X$ and $Y$, respectively, and $C_{XY}(\tau)$ is the cross-covariance function given by

$$C_{XY}(\tau) = E[(X(t-\tau) - \mu_X)(Y(t) - \mu_Y)] = E[X(t-\tau)Y(t)] - \mu_X\mu_Y. \tag{4}$$

Substituting equations (1) and (2) into equation (4), we get

$$
\begin{aligned}
C_{XY}(\tau) &= E[X(t-\tau)Y(t)] - \mu_X\mu_Y \\
&= E[X(t-\tau)aX(t-\tau_a) + X(t-\tau)W(t)] - \mu_X(a\mu_X + \mu_W) \\
&= E[aX(t')X(t' + \tau - \tau_a)] - a\mu_X^2 + E[X(t-\tau)W(t)] - \mu_X\mu_W \\
&= aC_{XX}(\tau - \tau_a) + C_{XW}(\tau) \\
&= aC_{XX}(\tau - \tau_a),
\end{aligned}
\tag{5}
$$

where we have assumed that $X$ and $W$ are independent.

Now substituting equation (1) into the denominator of equation (3), we get

$$
\begin{aligned}
\sigma_X\sigma_Y &= \sqrt{Var(X)Var(aX + W)} \\
&= \sqrt{Var(X)[Var(aX) + Var(W)]} \\
&= \sqrt{a^2\sigma_X^4 + \sigma_X^2\sigma_W^2} \\
&= \sigma_X\sqrt{a^2\sigma_X^2 + \sigma_W^2}.
\end{aligned}
\tag{6}
$$

Finally substituting equations (5) and (6) into equation (3), we get

$$
\begin{aligned}
r_{XY}(\tau) &= \frac{aC_{XX}(\tau - \tau_a)}{\sigma_X\sqrt{a^2\sigma_X^2 + \sigma_W^2}} \\
&= \sqrt{\frac{a^2\sigma_X^2}{a^2\sigma_X^2 + \sigma_W^2}}\left[\frac{C_{XX}(\tau - \tau_a)}{\sigma_X^2}\right] \\
&= \left(\sqrt{\frac{a^2\sigma_X^2}{a^2\sigma_X^2 + \sigma_W^2}}\right)r_{XX}(\tau - \tau_a) \\
&= Ar_{XX}(\tau - \tau_a),
\end{aligned}
\tag{7}
$$

where $r_{XX}(\tau)$ is the autocorrelation function of the driver $X$. That is, the cross correlation of the response and the driver is proportional to the time-shifted autocorrelation function of the driver, where the proportionality constant is $A$ and the time shift equals $\tau_a$. This relates to the familiar statement that the peak value of the square of the cross-correlation coefficient $[r_{XY}(\tau_a)]^2 = A^2 = a^2\sigma_X^2/(a^2\sigma_X^2 + \sigma_W^2)$ measures the proportion of the total variance explained by the directly driven signal.

Alternatively, we may write $A = \sqrt{S_a/(1 + S_a)}$ where $S_a = (a\sigma_X)^2/\sigma_W^2 = A/(1-A)$ is the signal-to-noise ratio of the first term of the right-hand side of equation (1) with respect to the second term. That is, the ratio of the directly driven part of the response, $aX(t - \tau_a)$, to the part that is independent of the driver, $W(t)$.

Now, generally, we would expect the signal-to-noise ratio $A$ and the time delay $\tau_a$ to vary from 1-month-long time series to another, in which case the mean cross-correlation coefficient from a number of month-long analyses is given by

$$\bar{r}_{XY}(\tau) = \frac{1}{N}\sum_{i=1}^{N} A_i r_{XX}(\tau - \tau_{ai}) \approx \frac{A}{N}\sum_{i=1}^{N} r_{XX}(\tau - \tau_{ai}), \tag{8}$$

where we have assumed that $A$ is given by the mean of $A_i$. This can be understood because for months with a similar $\tau_{ai}$ but different $A_i$, we expect the approximation in the right-hand side of equation (8) to hold for that $\tau_{ai}$. Also, we expect that if we were to divide the sum in equation (8) into different subgroups of similar $\tau_{ai}$, then the same approximation would hold for each. When computing equation (8), we pick $N = 1,000\tau_{ai}$ from the normal distribution $\mathcal{N}(\mu, \sigma)$ with mean $\mu$ and standard deviation $\sigma$ (values for these parameters are given in section 4). The models described in equations (7) and (8) are presented in section 4 to model the DPY response to IMF $B_y$ driving.

Now let us extend the model of equation (1) to include a second driver of the response

$$Y(t) = aX(t - \tau_a) + bX(t - \tau_b) + W(t).$$ (9)

In this case we have that the mean of Y is given by

$$
\begin{aligned}
\mu_Y &= E[Y] \\
&= E[aX(t - \tau_a) + bX(t - \tau_b) + W(t)] \\
&= E[aX(t - \tau_a)] + E[bX(t - \tau_b)] + E[W(t)] \\
&= aE[X(t - \tau_a)] + bE[X(t - \tau_b)] + E[W(t)] \\
&= a\mu_X + b\mu_X + \mu_W,
\end{aligned}
$$ (10)

the cross-covariance function becomes

$$
\begin{aligned}
C_{XY}(\tau) &= E[X(t - \tau)Y(t)] - \mu_X\mu_Y \\
&= E[X(t - \tau)aX(t - \tau_a) + X(t - \tau)bX(t - \tau_b) + X(t - \tau)W(t)] - \mu_X(a\mu_X + b\mu_X + \mu_W) \\
&= E[aX(t')X(t' + \tau - \tau_a)] - a\mu_X^2 + E[bX(t')X(t' + \tau - \tau_b)] - b\mu_X^2 + E[X(t - \tau)W(t)] - \mu_X\mu_W \\
&= aC_{XX}(\tau - \tau_a) + bC_{XX}(\tau - \tau_b) + C_{XW}(\tau) \\
&= aC_{XX}(\tau - \tau_a) + bC_{XX}(\tau - \tau_b),
\end{aligned}
$$ (11)

and for the denominator of equation 3, we get

$$
\begin{aligned}
\sigma_X\sigma_Y &= \sqrt{Var(X)Var(aX + bX + W)} \\
&= \sqrt{Var(X)[Var(aX) + Var(bX) + Var(W)]} \\
&= \sqrt{a^2\sigma_X^4 + b^2\sigma_X^4 + \sigma_X^2\sigma_W^2} \\
&= \sigma_X\sqrt{(a^2 + b^2)\sigma_X^2 + \sigma_W^2}.
\end{aligned}
$$ (12)

Finally substituting equations (11) and (12) into equation (3), we get

$$
\begin{aligned}
r_{XY}(\tau) &= \frac{aC_{XX}(\tau - \tau_a) + bC_{XX}(\tau - \tau_b)}{\sigma_X\sqrt{(a^2 + b^2)\sigma_X^2 + \sigma_W^2}} \\
&= \sqrt{\frac{a^2\sigma_X^2}{(a^2 + b^2)\sigma_X^2 + \sigma_W^2}}\left[\frac{C_{XX}(\tau - \tau_a)}{\sigma_X^2}\right] + \sqrt{\frac{b^2\sigma_X^2}{(a^2 + b^2)\sigma_X^2 + \sigma_W^2}}\left[\frac{C_{XX}(\tau - \tau_b)}{\sigma_X^2}\right] \\
&= \left(\sqrt{\frac{a^2\sigma_X^2}{(a^2 + b^2)\sigma_X^2 + \sigma_W^2}}\right)r_{XX}(\tau - \tau_a) + \left(\sqrt{\frac{b^2\sigma_X^2}{(a^2 + b^2)\sigma_X^2 + \sigma_W^2}}\right)r_{XX}(\tau - \tau_b) \\
&= Ar_{XX}(\tau - \tau_a) + Br_{XX}(\tau - \tau_b),
\end{aligned}
$$ (13)

and the mean cross-correlation coefficient from a number of month-long analyses is given by

$$
\begin{aligned}
\bar{r}_{XY}(\tau) &= \frac{1}{N}\sum_{i=1}^{N}A_i r_{XX}(\tau - \tau_{ai}) + \frac{1}{N}\sum_{i=1}^{N}B_i r_{XX}(\tau - \tau_{bi}) \\
&\approx \frac{A}{N}\sum_{i=1}^{N}r_{XX}(\tau - \tau_{ai}) + \frac{B}{N}\sum_{i=1}^{N}r_{XX}(\tau - \tau_{bi}).
\end{aligned}
$$ (14)

Thus, the cross-correlation coefficient of the response to a driver operating at two different lags is a weighted sum of the driver's autocorrelation function, which can be generalized to any number of lags. Multiresponse models based on equations (13) and (14) are presented in section 4 to model the DP2, DP1, and NBZ responses to IMF $B_z$ driving.

In what follows, the free parameters have been estimated by trial and error to fit the empirical cross-correlation function. Each component lag (i.e., $\tau_a$ and $\tau_b$) as first estimated (to about the 5-min resolution of the data) from the peaks in the empirical cross-correlation function. Corresponding amplitudes ($A$ and $B$) were chosen to give an approximate fit based on a fixed-$\tau$ model (equations (7) and (13)). The model fit was then improved by using the variable lag model (equations (8) and (14)). Here it was assumed that the mean lag $\mu$ was equal to $\tau$ of the fixed-$\tau$ model and then iterating the allowed variability $\sigma$ about the mean. A more sophisticated fitting method could have been used, but this seemed unnecessary based on the goodness of fit achieved.

## 4. Results

In Figures 2a–2d, we show the cross correlation between each month-long time series of a given equivalent current system and of a given IMF component, for lags varying from $-10$ to $+600$ min. Each gray line shows correlation with lag for a single monthly epoch. The thick red line shows the mean of these correlation curves taken over all months, illustrating the peak correlation for each equivalent current system. Figures 2e and 2j show the monthly and mean autocorrelations for IMF $B_y$ and $B_z$, respectively, in the same format as Figures 2a–2d. These mean IMF autocorrelations are used in one-component (equations (7) and (8)), two-component (equations (13) and (14)), and three-component models to reproduce the mean cross correlations from Figures 2a–2d. In the remainder of this section, we interpret the mean observed and modeled correlations for each equivalent current system in turn.

### 4.1. DPY Correlations

In Figure 2f, we compare the mean correlation between IMF $B_y$ and the DPY system (red line) with two models based on the IMF $B_y$ autocorrelation. The green line shows the modeled cross correlation between the driver and the response for a given single month, based on equation (2), using a fixed $A$ of $-0.68$ and a fixed $\tau_a$ of 20 min. The blue line shows the modeled mean cross correlation based on equation (3), using the same fixed $A$ and a variable $\tau_a$, drawn from a normal distribution with $\mu = 20$ min and $\sigma = 10$ min. The latter model fits the data well, and we thus conclude that about $A^2 * 100 = 46\%$ of the variance of the EOF amplitude time series from the DPY cluster is explained by IMF $B_y$. We infer that the response at $\tau_a = 20$ min is the DPY response to magnetopause reconnection. This imparts an azimuthal stress on newly reconnected field lines, creating an azimuthal circulation of plasma, and the corresponding antiparallel circulation of equivalent current shown in Figure 1a (e.g., Friis-Christensen & Wilhjelm, 1975; Tenfjord et al., 2015). The negative sense of the correlation in Figures 2a and 2f is in agreement with the expectation that duskward (positive) IMF $B_y$ causes an eastward DPY circulation, since the associated pattern in Figure 1a shows a westward circulation.

### 4.2. DP2 Correlations

In Figure 2g, we compare the mean correlation between IMF $B_z$ and the DP2 system with two models based on the IMF $B_z$ autocorrelation. The first model (green line) is a two-component model which assumes that the DP2 response is a weighted superposition of IMF $B_z$ at two different lags. Hence, the cross correlation is the weighted sum of two evaluations of equation (2), one using a fixed $A_1$ of $-0.25$ and a fixed $\tau_1$ of 20 min and another using a fixed $A_2$ of $-0.53$ and a fixed $\tau_2$ 60 min. The second model (blue line) is a superposition of two evaluations of equation (3), one with $A_1 = -0.25$, $\mu_1 = 20$ min, and $\sigma_1 = 10$ min and the other with $A_2 = -0.53$, $\mu_2 = 60$ min, and $\sigma_2 = 30$ min. The latter model fits the data well, and thus, we conclude that $A_1^2 * 100 = 6.25\%$ of the total variance is described by the first component of the response, $A_2^2 * 100 = 28.09\%$ of the total variance is described by the second component, and $(A_1^2 + A_2^2) * 100 = 34.34\%$ of the total variance is described by the two components combined (i.e., by IMF $B_z$). We attribute the first of the two responses at 20-min lag to magnetopause reconnection, the same physical source which describes the DPY group. The second response at 60-min lag describes the response to nightside reconnection. Together, the two responses create the two-cell convection pattern shown in Figure 1b (Nishida, 1968a), expected from the expanding-contracting polar cap model (Lockwood et al., 1990; Lockwood & Cowley, 1992). We could fit the data equally well with a more parsimonious single response model ($A = -0.79$, $\mu = 40$ min, and $\sigma = 35$ min), but since DP2 is driven by both dayside and nightside reconnections, we have a sound physical motivation to favor the two-component model.

### 4.3. NBZ Correlations

In Figure 2h, we compare the mean correlation between IMF $B_z$ and the NBZ system with two models based on the IMF $B_z$ autocorrelation. The first model (green line) is a two-component model based on two evaluations of equation (2), one using a fixed $A_1$ of 0.58 and a fixed $\tau_1$ of 15 min and another using a fixed $A_2$ of $-0.63$ and a fixed $\tau_2$ of 90 min. The second model (blue line) is a superposition of two evaluations of equation (3), one with $A_1 = 0.58$, $\mu_1 = 15$ min, and $\sigma_1 = 5$ min and the other with $A_2 = -0.63$, $\mu_2 = 90$ min, and $\sigma_2 = 40$ min. This model fits the data well, and we conclude that $A_1^2 * 100 = 33.64\%$ of the total variance is described by the first component of the response, $A_2^2 * 100 = 39.69\%$ of the total variance is described by the second component, and $(A_1^2 + A_2^2) * 100 = 73.33\%$ of the total variance is described by the two components combined (i.e., by IMF $B_z$). Again, we infer that the first of the two responses at 15-min lag is driven by magnetopause reconnection. Given the spatial sense of the NBZ pattern shown in Figure 1c, we expect the observed positive correlation between this pattern and the IMF $B_z$ component. It is likely that this pattern has a slightly faster directly driven response to magnetopause reconnection than observed for DPY or DP2 because NBZ is physically smaller and thus takes less time to reconfigure. The second of the two-component responses at 90-min lag implies that some time after the IMF is directed northward, it will turn southward. At this point the solar wind will drive an ionospheric two-cell convection pattern which is (somewhat) opposite to the NBZ pattern in Figure 1c, and hence, the cross correlation of NBZ and IMF $B_z$ can be expected to reverse in sign. The broad peak of the secondary response reflects the very variable timescale for the IMF $B_z$ component to reverse in sign following an initial northward state.

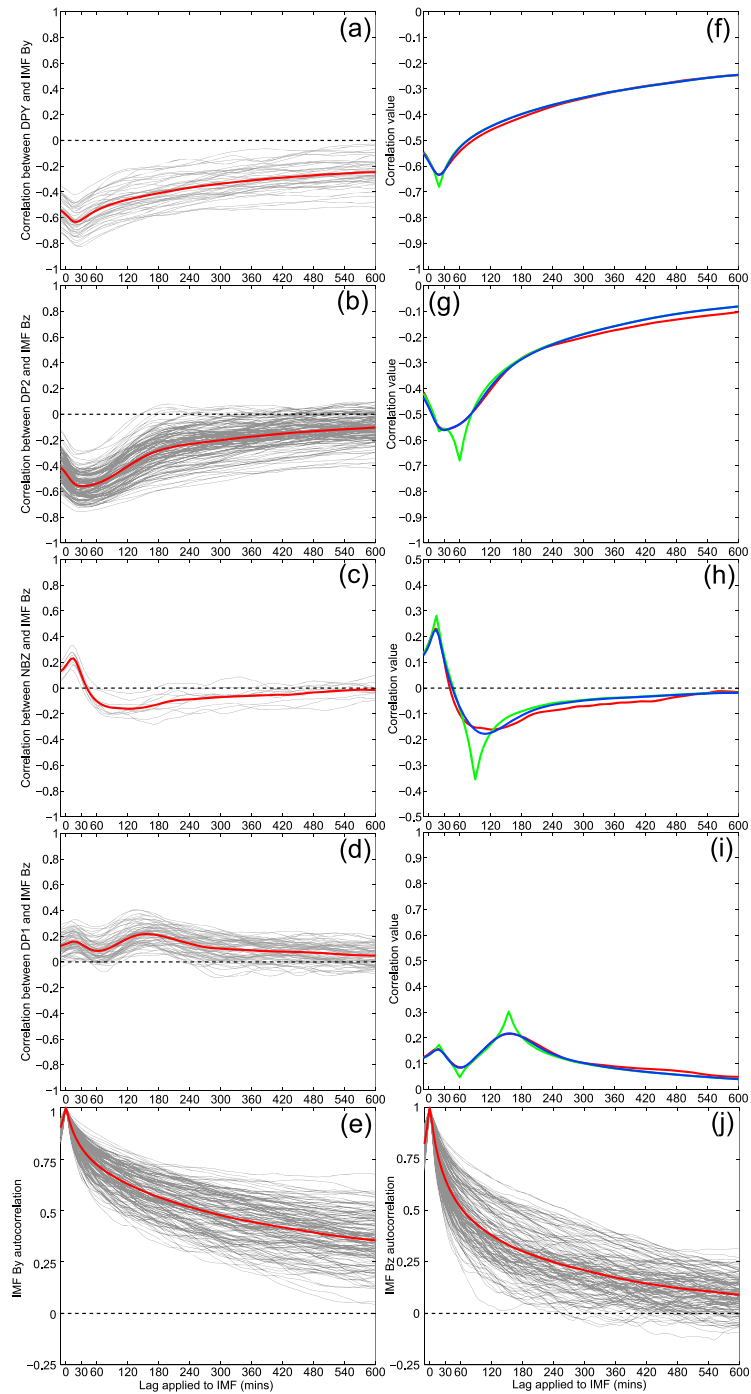

**Figure 2.** (a–d) Monthly correlations between interplanetary magnetic field (IMF) $B_y$ or $B_z$ and the time series of amplitudes for each of the equivalent current systems DPY (a), DP2 (b), NBZ (c), and DP1 (d). Each gray line shows correlation with lag for a single monthly epoch. The red line shows the mean correlation series, taken across all monthly epochs for each equivalent current system. The dashed black line is at zero correlation. (e and j) Monthly autocorrelations of the IMF $B_y$ (e) and $B_z$ (j) components as a function of lag. Line colors are the same as (a)–(d). (f–i) The red line is the same as that in the corresponding adjacent panel from (a)–(d), that is, (a) and (f) form a pair. The green line is the modeled correlation series from equation (2), and the blue line is given by equation (3).

### 4.4. DP1 Correlations

In Figure 2i, we compare the mean correlation between IMF $B_z$ and the DP1 system with two models based on the IMF $B_z$ autocorrelation. The first model (green line) is now a three-component model based on equation (2), one with fixed $A_1$ of 0.17 and a fixed $\tau_1$ of 20 min, another with fixed $A_2$ of $-0.19$ and a fixed $\tau_2$ of 60 min, and the last with fixed $A_3$ of 0.32 and a fixed $\tau_3$ of 155 min. The second model (blue line) is a superposition of three evaluations of equation (3), one with $A_1 = 0.17$, $\mu_1 = 20$ min, and $\sigma_1 = 8$ min, another with $A_2 = -0.19$, $\mu_2 = 60$ min, and $\sigma_2 = 15$ min, and the last with $A_3 = 0.32$, $\mu_3 = 155$ min, and $\sigma_3 = 30$ min. This model fits the data well, and we conclude that $A_1^2 * 100 = 2.89\%$ of the total variance is described by the first component of the response, $A_2^2 * 100 = 3.61\%$ of the total variance is described by the second component, $A_3^2 * 100 = 10.24\%$ of the total variance is described by the third component, and $(A_1^2 + A_2^2 + A_3^2) * 100 = 16.74\%$ of the total variance is described by the three components combined (i.e., by IMF $B_z$). We infer that the first of these three responses (at 20 min) is actually a prompt DP2 response to magnetopause reconnection. The DP1 is not perfectly isolated in the EOF modes, and the DP1 cluster still contains some DP2 signal. For instance, away from the expected location of the DP1 westward electrojet between 21 MLT and 03 MLT, the SEIMF in the auroral electrojet and polar cap regions in Figure 1d is spatially similar to that in Figure 1b yet reversed in sign (i.e., the spatial pattern here resembles a negative DP2 system). This sign reversal is why the prompt response in Figure 2i is a positive correlation, while the prompt response in Figure 2g is a negative correlation. The second modeled response is the substorm—the more-negative correlation at 60-min lag is consistent with the DP1 westward electrojet forming at substorm onset, following a growth phase lasting approximately 1 hr after an IMF $B_z$ southward turning. The fact that the cross correlation at 60-min lag is not actually negative reflects the overall dominance of the inclusion of signal from the stronger DP2 pattern. Lastly, the third response possibly reflects the tendency of the substorm to repeat on a 2-hr timescale. The first substorm enhances ionospheric conductivity due to associated particle precipitation, and so the variance of the response in the second substorm is larger for a given IMF input.

## 5. Discussion

The most predictable modes from a given single IMF component are DPY (46% of variance explained) and NBZ (73% explained). While the NBZ modes would appear to be weakly related to the IMF (peak correlation $\sim 0.2$), we find that this is a superposition of two opposing responses at different lag, each more strongly related to the IMF. DP1 is the least well-explained mode set (at 17%) based on the IMF, which is expected since it represents the equivalent current response to substorm onset. This is an impulsive response to a time-integrated property of the IMF and arguably (Freeman & Morley, 2009; Morley et al., 2007) also to an additional trigger from upstream IMF variations in around half of cases (Milan et al., 2007). It is likely that more of the variability of the $B_z$-controlled modes would be explained by a nonlinear function of the solar wind parameters (e.g., Boynton et al., 2011; Finch & Lockwood, 2007; Newell et al., 2007; Spencer et al., 2011), but we do not investigate this here.

All four mode sets exhibit a prompt response to magnetopause reconnection. This is the well-known directly driven response of the electrojets to the IMF (Kamide & Kokubun, 1995). At longer lags, we find that the DP2 and DP1 responses are more strongly correlated to the IMF variations than the initial response, despite being presumably less directly driven (e.g., Milan et al., 2018; Weimer, 2013). This seems counterintuitive but can be explained via an ionospheric conductivity feedback. When magnetopause reconnection associated with the prompt response to the IMF is sustained for tens of minutes, a substorm will likely ensue that will increase particle precipitation and conductivity in the auroral zone. After this, a unit change in the driver will cause a larger response in the SEIMF there; hence, the IMF explains more of the SEIMF variability for DP2 and DP1 perturbations at longer lags. Despite their stronger correlation with the IMF, we do not imply that the long lag variations are more predictable. Rather, we think that the long lag contributions to the SEIMF explain more of the total variance because more of the variance comes from after substorm onset, due to the higher post-onset conductivity. In addition, the variation in the monthly lag is much larger than for the prompt responses.

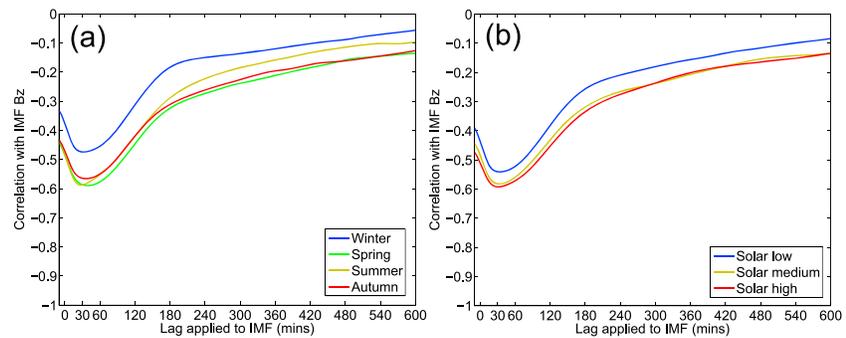

**Figure 3.** Correlations between interplanetary magnetic field (IMF) $B_z$ and the time series of amplitudes for the DP2 equivalent current system, grouped by (a) season and (b) solar cycle phase according to $F_{10.7}$. Only the means of the grouped data are shown.

Evidence to support the link between ionospheric conductivity and the mode response to the IMF can be provided by taking subsets of the correlations according to season and solar cycle phase. This cannot be performed for the DPY, DP1, or NBZ modes, since each is absent from the EOF analysis in some seasons. However, we can assess how the relationship between the DP2 modes and the IMF $B_z$ component varies according to season and solar cycle, which we show in Figure 3. Figure 3a shows the data from Figure 2b, grouped by season. Here we define season as pairs of adjacent months with a single-month gap in-between. The year is divided as follows: summer (June and July); autumn (September and October); winter (December and January); and spring (March and April). In Figure 3b, we have divided the data from Figure 2b by solar cycle phase, where solar low is defined from monthly $F_{10.7}$ values below 130 fu, solar medium is between 130 and 170 sfu, and solar maximum is greater than 170 sfu. We find that seasons and phases associated with low ionospheric conductivity show lower mean correlation between the DP2 modes and IMF $B_z$ for both prompt and indirect responses. Conversely, in summer, when the conductivity is the highest (Figure 3a), we see a small relative increase in the proportion of the prompt response explained by IMF $B_z$, consistent with the conductivity perturbations caused by prior IMF variations being a smaller part of the total conductivity then.

There are a number of other studies which have looked specifically at how the magnetosphere-ionosphere system responds to the solar wind as a function of lag time. Here we discuss the results we have presented in the context of these studies.

Other authors have computed EOF analyses (or mathematic equivalents) of geophysical data from the polar regions and correlated the discovered modes with the IMF as a function of lag (Baker et al., 2003; Milan et al., 2015; Sun et al., 1998). Baker et al. used auroral ultraviolet brightness and discovered somewhat different modes to those of Shore et al. (2018), making direct comparison with the results of our study difficult. However, it is clear that the correlations of the ultraviolet modes with the IMF have their peaks at different (typically longer) characteristic lags than the modes assessed in this study, with ranges from 5 to 180 min.

The study of Sun et al. (1998) is closest to ours, since their modes are also discovered from ground-based magnetic data. The authors analyzed the modes from 2 days of data in spring equinox and discovered similar relationships between the DP2 and DP1 modes and the IMF as we have reported here. Sun et al. also indicated that DP2 and DP1 were imperfectly separated in the mode decomposition, as we have found here too. The advances of our study are that we have assessed more—and different—current systems (i.e., DPY and NBZ) over the span of a solar cycle, and we have quantified the proportion of each mode explained by the IMF, for responses at varying time delays.

Browett et al. (2017) have correlated the IMF $B_y$ measurements with $B_y$ measured in the terrestrial magnetotail and assessed this relationship as a function of lag time. The authors found multiple peaks in this correlation, at lags commonly larger than 1 hr, arising from reconnection in an asymmetrically loaded magnetotail. We find no systematic (i.e., present in several months) correlation peaks in the relationship between DPY and IMF $B_y$ (shown in Figure 2a) at lags greater than 20 min. However, we note that Browett et al. (2017) found a pronounced control from solar wind speed and IMF $B_z$ sign on the timescale for correlation between IMF $B_y$ and magnetotail-measured $B_y$. We have not accounted for this behavior here, and nightside reconnection may account for the deviation between the red and blue lines in Figure 2f between lags of 60 and 240 min. It is possible that subdividing the Shore et al. (2018) data set could reveal a coherent DPY-$B_y$ correlation at long lag. Our results for the DPY prompt response are consistent with the study of Tenfjord et al. (2015), which showed (their Figure 2) that $B_y$ stress is exerted on the ionosphere for 5–20 min after reconnection, following which time a "stable" hemispheric asymmetry is attained.

Lastly, we wish to highlight the study of Maggiolo et al. (2017) who have performed a correlation between geomagnetic indices and solar wind parameters spanning 2000–2010, also as a function of lag. The authors' most pertinent finding was that of a peak in the correlation between IMF $B_z$ and the AE index at a lag of 35 min. Maggiolo et al. interpreted this delay as a particularly rapid substorm growth phase—we disagree with this interpretation. Instead, we suggest that the 35-min delay between IMF $B_z$ and the AE index is due to the superposition of the two responses of the DP2 system to IMF $B_z$ which we found in section 4.2.

Our study has resolved the proportion of the geomagnetic variation which can be described from a given IMF component variation and how this varies with spatial location and time delay. This localized temporal behavior is not accounted for in otherwise state-of-the-art models of space weather, for instance, by Weimer (2013) and Laundal et al. (2018). Both of these models are based on a regression of spherical harmonic coefficient amplitudes onto an ensemble of solar wind parameters. This technique confers two main shortcomings. First, the same time lag in the response to solar wind driving must be assumed for all locations, because the spherical harmonics are defined globally. The models apply temporal averaging to mitigate this assumption, which reduces the temporal precision of their predictions. The Weimer (2013) model applies 25-min smoothing, and the AMPS model (Laundal et al., 2018) applies 20-min smoothing. Second, neither of these models separates the expected space weather response by magnetic variation type (i.e., by equivalent current system). Our technique here separates the responses of different current systems, and so we do not require temporal smoothing when describing the localized response. From a practical point of view, our new findings identify the ionospheric regions which can be best (and worst) predicted from the solar wind. This new information is crucial to understanding and empirically predicting the space weather impacts of the connected Sun-Earth environment on Earth systems and infrastructure.

## 6. Conclusions

We use the EOF reanalysis of ground-based magnetic field perturbations, provided by Shore et al. (2018), to investigate the response of the ionospheric equivalent current modes DPY, DP2, NBZ, and DP1 to variations in IMF $B_y$ and $B_z$. We quantify the different temporal latencies in these responses and the proportion of the variance of the modes which is described by the IMF. Thus, we resolve the predictability of the different equivalent current systems in the context of solar wind conditions.

We find that IMF $B_y$ accounts for 46% of the DPY mode variability and that IMF $B_z$ accounts for, respectively, 34%, 73%, and 17% of the DP2, NBZ, and DP1 modes. Each of these four sets of modes exhibits a prompt response to magnetopause reconnection at 15- to 20-min lag, following a perturbation at the Earth's bow shock nose. The DP2 and DP1 modes show a secondary response driven by magnetotail reconnection at 60-min lag. In addition, the DP1 modes exhibit a third response at 155 min, which may reflect the most common substorm repeat interval. The NBZ modes exhibit a secondary response at 90 min, which may reflect the timescale for the IMF to persist in a northward state. For the DP2 and DP1 modes, we find that the IMF explains more of the variation of the secondary and tertiary responses than the initial prompt response. This is consistent with substorm-enhanced ionospheric conductivity increasing the equivalent current response amplitude for a unit input from the IMF driver.

We confirm the supposition (introduced by Shore et al., 2018) that using graph theory to group geomagnetic EOF modes into similar patterns can resolve the disturbance-polar equivalent current systems identified from historic ground magnetic records. We show that these groups of modes each correlate differently with the IMF driving in terms of the timescale and amplitude of the response, and we quantify the modulation of the DP2 response according to solar cycle and season. Our results are consistent with the spatial patterns of the modes and our expectations from established theory. Our improved description of the IMF driving offers a better understanding of the underlying physics and timescales. This should improve future estimates of solar wind-magnetosphere coupling and of the geoeffectiveness of solar activity, both of which are challenging issues for describing space weather.

**Acknowledgments**

This work was funded by the Natural Environment Research Council under grants NE/J020796/1 and NE/N01099X/1. We would like to thank Eelco Doornbos, Steve Browett, Maria-Theresia Walach, John Coxon, Suzie Imber, Gareth Chisham, and three anonymous reviewers for discussions which improved the paper. The EOF reanalysis model used in this study is available in the supporting information of Shore et al. (2018). OMNI data were downloaded from ftp://spdf.gsfc.nasa.gov/pub/data/omni/high_res_omni/monthly_1min/ on 06 March 2014. $F_{10.7}$ data were obtained from ftp://ftp.ngdc.noaa.gov/STP/GEOMAGNETIC_DATA/INDICES/KP_AP/ on 19 December 2012. For the SuperMAG ground magnetometer data, we gratefully acknowledge: INTERMAGNET (we thank the national institutes that support its contributing magnetic observatories and INTERMAGNET for promoting high standards of magnetic observatory practice; www.intermagnet.org); USGS, Jeffrey J. Love; CARISMA, PI Ian Mann; CANMOS; The S-RAMP Database, K. Yumoto and K. Shiokawa; The SPIDR database; AARI, Oleg Troshichev; The MACCS program, M. Engebretson, Geomagnetism Unit of the Geological Survey of Canada; GIMA; MEASURE, UCLA IGPP and Florida Institute of Technology; SAMBA, Eftyhia Zesta; 210 Chain, K. Yumoto; SAMNET, Farideh Honary; the institutes who maintain the IMAGE magnetometer array, Eija Tanskanen; PENGUIN; AUTUMN, Martin Connors; DTU Space, Anna Naemi Willer; South Pole and McMurdo Magnetometer, Louis J. Lanzarotti and Alan T. Weatherwax; ICESTAR; RAPIDMAG; PENGUIn; British Antarctic Survey; McMac, Peter Chi; BGS, Susan Macmillan; Pushkov Institute of Terrestrial Magnetism, Ionosphere and Radio Wave Propagation (IZMIRAN); GFZ, Juergen Matzka; MFGI, B. Heilig; IGFPAS, J. Reda; University of L'Aquila, M. Vellante; and SuperMAG, Jesper W. Gjerloev.